\documentclass[aps,prd,showpacs,showkeys,superscriptaddress,twocolumn]{revtex4-1}
\usepackage[colorlinks=true,linkcolor=blue,anchorcolor=blue,citecolor=blue,urlcolor=blue,breaklinks=true]{hyperref}
\usepackage{graphicx}
\usepackage{amsmath}
\usepackage{amssymb}
\usepackage{slashed}
\usepackage{latexsym}
\usepackage{epsfig}
\usepackage{amsbsy}
\usepackage{array}
\usepackage{amssymb}
\usepackage{setspace}
\usepackage{bm}
\usepackage{lipsum}
\usepackage{mathrsfs}
\usepackage{float}
\usepackage{color}
\usepackage[T1]{fontenc}
\usepackage{mathptmx}
\DeclareMathAlphabet{\mathcal}{OMS}{cmsy}{m}{n}
\DeclareSymbolFont{largesymbols}{OMX}{cmex}{m}{n}

\begin{document}
\author{Peng Cheng} \email{pengC@smail.nju.edu.cn} \affiliation{Department of Physics, Nanjing University, Nanjing 210093, China}
\author{Xiaofeng Luo} \email{xfluo@mail.ccnu.edu.cn} \affiliation{Key Laboratory of Quark $\&$ Lepton Physics (MOE) and Institute of Particle Physics, Central China Normal University, Wuhan 430079, China}
\author{Jialun Ping} \email{jlping@njnu.edu.cn} \affiliation{Department of Physics, Nanjing Normal University, Nanjing 210023, China}
\author{Hongshi Zong} \email{zonghs@nju.edu.cn}
\affiliation{Department of Physics, Nanjing University, Nanjing 210093, China}
\affiliation{Joint Center for Particle, Nuclear Physics and Cosmology, Nanjing 210093, China}
\affiliation{Nanjing Proton Source Research and Design Center, Nanjing 210093, China}

\title{Finite volume effects on quarkonium dissociation temperature in an impenetrable QGP sphere}

\begin{abstract}
The system of a quarkonium confined by an impenetrable spherical cavity filled with a hot
quantum chromodynamics (QCD) medium is studied by solving the Schr\"{o}dinger equation.
This is the first time this issue has been raised for discussion. The Schr\"{o}dinger equation
with an appropriate boundary condition of a quarkonium in an impenetrable cavity filled with a hot medium
is derived. The numerical results are obtained with the help of Gaussian Expansion Method.
Binding energies and radii of the ground and low-excited states are obtained as a function of the medium
temperature and the cavity radius. We find the behaviour of quarkonium in this cavity is different
from that in infinite space. Our results show that the quarkonium dissociation temperature decreases as the
cavity radius decreases and the finite volume effects on the ground state are more obvious than on the excited states.
We also find that the less mass of the constituents and the bigger radius of the quarkonium lead the finite
volume effects to become more obvious.

\bigskip

\noindent Keywords: Quarkonium; Dissociation temperature; Gaussian Expansion Method; Finite volume effects.
\end{abstract}

\pacs{12.38.Mh, 12.39.-x, 25.75.Nq}

\maketitle

\section{Introduction}
The quantum chromodynamics (QCD) predicts that at high temperature and/or density there is a phase transition from hadron
to quark gluon plasma (QGP). The QGP is generally believed to be produced during the relativistic heavy ion collisions.
In statistical QCD, deconfinement and the properties of the resulting QGP can be investigated by studying the behavior of
heavy quark bound states in a hot medium \cite{Matsui:1986dk}. In 1986, Matsui and Satz pointed out the suppression of $J/\psi$
can be recognized as a signal of QGP formation in the relativistic heavy ion collisions \cite{Matsui:1986dk}. Since then,
the suppression of quarkonium production in a QGP has been a subject with intensive interest.
There have been many theoretical researches \cite{Satz:1988rw,Kharzeev:1996se,Vogt:1999cu,Thews:2000rj,Karsch:2005nk,Laine:2006ns,Beraudo:2007ky,PhysRevC.92.061901,
Du:2017hss,PhysRevC.97.014908,Blaizot:2017ypk,Yao:2018sgn,Du:2018wsj,PhysRevD.99.096028}
and experimental studies at Super Proton Synchrotron (SPS) \cite{Gonin:1996wn,Abreu:2000ni,Quintans:2006wc,Alessandro:2006dc},
Relativistic Heavy Ion Collider (RHIC) \cite{Adare:2006ns,Abelev:2009qaa,Adamczyk:2013tvk} and
Large Hadron Collider (LHC) \cite{Khachatryan:2016xxp,Santos:2019fyo,Paul:2019nws}.
In the past thirty years, the work on quarkonium dissociation temperatures has also attracted great
interest \cite{Karsch:1987pv,Satz:2000bn,Digal:2001iu,Satz:2005hx,Karsch:2005nk,Qu:2012zz,Cheng:2018pwl},
because they are related to the suppression of quarkonium production. All these theoretical studies
calculate the dissociation temperatures of quarkonium in an infinite space, while the quarkonium produced by
relativistic heavy ion collisions is actually in a finite-size fireball
\cite{PhysRevC.16.1493,Heinz:2002gs,Zhao:2011cv,Datta:2014wga,Liu:2014rsa,Escobedo:2016iuu,PhysRevD.97.074009,PhysRevC.99.054901},
formed by the relativistic heavy ion collisions. Recently, researchers have begun to notice the effects on
quarkonium production arising from the volume of fireball \cite{Liu:2014rsa,PhysRevC.99.054901}. They discussed
the finite volume effects on the suppression of quarkonia at early time after the collision.

At high temperature and/or density, the interactions between the heavy quark and antiquark pairs are screened \cite{Digal:2005ht}
and the binding energy will decrease. As a result, the heavy quark bound states will start to dissociate when
the binding energy becomes low enough (and its radius becomes large enough).
In Refs. \cite{Karsch:2005nk,Satz:2000bn,Digal:2001iu,Satz:2005hx,Qu:2012zz}, the dissociation of quarkonium has been studied
in the Schr\"odinger equation formalism. In our previous work \cite{Cheng:2018pwl}, we also calculate the dissociation
temperature of quarkonium by solving Schr\"{o}dinger equation with the help of Gaussian Expansion Method (GEM),
an efficient and powerful method for few-body system \cite{Hiyama:2003cu}. All of these works are based on an assumption
that quarkonium lies in an infinitely large medium, while the volume of the hot medium (QGP) is finite in experiment,
especially at early time after the collision \cite{PhysRevC.99.054901}. The volume of QGP is about at the same scales
as that of a nucleus. In order to understand the results of relativistic heavy ion collision experiments better,
considering the effect of the finite-size fireball volume on the dissociation temperature of the heavy quark bound
states is necessary. In this work, we will study the finite volume effects on quarkonium dissociation temperature
based on the previous work \cite{Cheng:2018pwl}.

In our previous work, the temperature-dependent potential between the heavy quark and antiquark was obtained by fitting
the free energy of a heavy quark-antiquark system $F_{Q\bar Q}(r,T)$ which can be calculated in lattice QCD
\cite{Kaczmarek:2002mc,Kaczmarek:2005uv}. The analytical form of $F_{Q\bar Q}(r,T)$ was constructed based on the
Debye-H\"uckel theory \cite{Dixit:1989vq}, and its temperature-dependent parameters were determined by fitting the lattice data.
To study the finite volume effects on quarkonium dissociation temperature, we present a simplified model that the quarkonium produced in
the relativistic heavy ion collisions is treated as being confined in an impenetrable spherical cavity filled with a hot QCD medium.
The reason why our model is a a simplified model is that the real fireball produce in heavy ion collisions
is not impenetrable and particles (such as unbound heavy quarks) that reach the boundary of the QGP just hadronize and fly out of the QGP, as hadrons.
This real case is difficult to solve. But our model can help us to get a first insight to the finite volume effects on quarkonium dissociation temperature.
In fact, this model involves a fundamental question of quantum mechanics, namely how to solve a bound state constrained in a
finite space region. Since $1937$, studies on the properties of a hydrogen atom confined in the impenetrable spherical cavity
have received much attention. It was first investigated by Michels $et\ al.$ \cite{michels1937remarks}, followed by many
authors \cite{1938AnP...424...56S,1956CaJPh..34..914H,1984JChPh..80.1569A,1991AmJPh..59..931M,2002JPhB...35..255P,Kang2013}.
And the work was extended to a helium atom confined in the impenetrable spherical cavity \cite{wen2006calculation}.
Recently, the authors of
Ref. \cite{2019arXiv190205355P} proposed a new model for hydrogen atom by solving Schr\"{o}dinger equation with a correct boundary condition.
In this paper, with the help of this new model, we will study the finite volume effects on quarkonium dissociation temperature.
The numerical results are obtained with the help of GEM,
whose validity and reliability on calculating dissociation temperature has been verified in our previous work \cite{Cheng:2018pwl}.
By solving Schr\"{o}dinger equation, we obtain the temperature dependence of binding energy and radius for the ground and low-lying
excited states. For infinite space, we usually use the binding energy and/or radius to define the dissociation temperature and
the dissociation temperature is the point where the binding energy decreases to zero and the radius increases to infinite.
However, the quarkonium here is confined in an impenetrable cavity and its radius is impossible to become infinite.
So we determine the dissociation temperature according to the binding energy, rather than the radius.

This paper is organized as follows: In Sec. \ref{sec:model}, we explain our model in detail and the corresponding
non-relativistic Hamiltonian is presented. In Sec. \ref{sec:method}, the method is explained.
In Sec. \ref{sec:res}, we show the numerical results. Sec. \ref{sec:summ} contains discussions and conclusions.

\section{The Model}\label{sec:model}

Due to the large mass of heavy quarks, the non-relativistic potential model is successfully applied to the study of
charmonium and bottomonium states. The most frequently used potential for a $Q\bar Q$ system is the Cornell potential
\begin{equation}
    V_{Q \bar Q}(r) = -\frac{\alpha}{r} + \sigma r,
\label{eq:cornel}
\end{equation}
where $\alpha$ is the gauge coupling constant and $\sigma$ is the string tension. In this equation, the first term
corresponds to the Coulomb interaction between static charges and the second term is due to the formation of flux tube
or string between the quark and the antiquark when they are pulled apart.
Substituting the potential into the Schr\"odinger equation of quarkonium, we can determine the potential parameters
$\alpha$ and $\sigma$, the charm quark mass $m_c$ and the bottom quark mass $m_b$ by fitting the spectroscopy of quarkonium.
According to the work in Ref. \cite{Satz:2005hx}, these parameters are listed in Table \ref{tab:para}
\begin{table}[!h]
\begin{center}
\caption{\label{tab:para} Parameters in the potential model and quark mass.}
\renewcommand\arraystretch{1.8}
\begin{tabular}{ccccc}
\hline
  & $m_c$[GeV] & $m_b$[GeV] & $\alpha$ & $\sqrt {\sigma}$[GeV] \\
\hline
Ref. \cite{Satz:2005hx} & 1.25 & 4.65 & $\frac{\pi}{12}$ & 0.445 \\
\hline
\end{tabular}
\end{center}
\end{table}

At high temperature and/or density, the interaction between the constituents of quarkonium is screened. We
shall consider the case of vanishing baryon-number density (baryons and antibaryons in equal numbers). In previous work,
the temperature dependent potential has been extracted from the free energy of a heavy quark-antiquark system
$F_{Q\bar Q}(r,T)$ which is calculated in lattice QCD \cite{Kaczmarek:2002mc,PhysRevD.71.114510}. The analytical
form of $F_{Q\bar Q}(r,T)$ can be obtained based on studies of screening in Debye-H\"uckel theory.
It is \cite{Digal:2005ht}
\begin{eqnarray}
F_{Q\bar Q}(r,T)& = & -\frac{\alpha}{r}\left[e^{-\mu r}+\mu r\right]+\frac{\sigma}{\mu}\left[
    \frac{\Gamma\left(1/4\right)}{2^{3/2} \Gamma\left(3/4\right)}  \right.    \nonumber \\
 & & \left.  -\frac{\sqrt{\mu r}}{2^{3/4}\Gamma\left(3/4\right)}K_{1/4}\left[\left(\mu r\right)^2
  +\kappa \left(\mu r\right)^4\right] \right],
    \label{eq:free}
\end{eqnarray}
where the screening mass $\mu$ and the parameter $\kappa$ are temperature-dependent, and $K_{1/4}[x]$ is the modified
Bessel function. The T-dependent $\mu$ and $\kappa$ can be determined by fitting $F_{Q\bar{Q}}(r,T)$ to the
lattice results obtained in 2-flavor QCD \cite{PhysRevD.71.114510}.
In Ref. \cite{Digal:2005ht}, the authors obtained the fitting results for the temperature dependence of $\mu(T)$ and $\kappa(T)$
and showed the fitting curves together with the lattice results. Their results showed that
the analytical form of $F_{Q\bar Q}(r,T)$ fitted the lattice data quite well for all $r$ and in a broad range of
temperatures from $0.8T_c$ to $2T_c$. According to the argument in Ref. \cite{PhysRevD.98.116010}, we assume that
the interquark potential is just the internal energy, i.e., $V=F+ST$ where $S$ is the entropy $S=-\partial F/\partial T$
in our model. So the potential between quark and antiquark in a hot QCD medium is written as
\begin{equation}
    V_{Q\bar Q}(r,T) = F_{Q\bar Q}(r,T) - T \frac{\partial F_{Q\bar Q}(r,T)}{\partial T}.
\label{eq:interquark}
\end{equation}
Then we obtain the mass of charmonium (or bottomonium) state $i$, i.e. $M_i$, at temperature $T$ by solving
the Schr\"odinger equation
\begin{equation}
\left[\sum_{j=1}^2 (\frac{\boldsymbol{p}_j^2}{2m_j}+m_j) - T_{cm} + V_{Q \bar Q}(r,T)\right]\Psi_i=M_i \Psi_i,
\label{eq:mass11}
\end{equation}
where $m_j$ is the constituent quark mass of the $j$-th quark and $T_{cm}$ is the center-of-mass kinetic energy.
$\boldsymbol{r}=\boldsymbol{r}_1-\boldsymbol{r}_2$ is the relative motion coordinate.
We define the binding energy of charmonium (or bottomonium) state $i$ as
\begin{equation}
\Delta E_i(T)=-\epsilon_i(T)=-(M_i-2m_{Q}-V_{Q\bar Q}(\infty,T)).
\label{eq:bindingenergy1}
\end{equation}
Combining Eq. (\ref{eq:mass11}) with Eq. (\ref{eq:bindingenergy1}), we can obtain
\begin{equation}
\left[\sum_{j=1}^2 \frac{\boldsymbol{p}_j^2}{2m_j} - T_{cm} + V_{Q \bar Q}(r,T)-V_{Q \bar Q}(\infty,T)\right]\Psi_i=\epsilon_i(T) \Psi_i.
\label{eq:schr1}
\end{equation}
Solving this Schr\"odinger equation, we obtain the binding energy $\Delta E_i(T)$ and the
corresponding wave function at temperature $T$. We define the radius as
\begin{equation}
\sqrt{\langle r^2 \rangle}=\left[\int \Psi^* r^2 \Psi d\tau\right]^{\frac 12}
\label{eq:radii1}
\end{equation}
where the $d\tau$ represents the volume element of the integral, i.e. $d^3 \boldsymbol{r}$.
Then we can use the resulting wave function to calculate the $T$-dependent radius. When the binding energy
vanishes, the bound state $i$ no longer exists. So $\Delta E_i(T)=0$ determines the dissociation temperature
for state $i$. This is what we have done in Ref. \cite{Cheng:2018pwl}. The model mentioned above is just suitable
for the case that the hot QCD medium is infinitely large. If the volume of the hot medium (QGP) is finite,
We have to modify this model.

\begin{figure}[!htbp]
\centering
\includegraphics[width=0.35\textwidth]{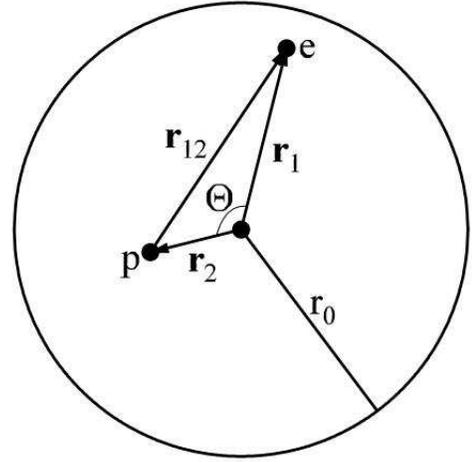}
\caption{\label{fig:hydrogen} A hydrogen in an impenetrable spherical cavity\cite{2019arXiv190205355P}.}
\end{figure}

Here, we present a simplified model. Taking the finite volume of a hot medium into account, the model we are considering
is described as the quarkonium confined in an impenetrable spherical cavity which is filled with a hot QCD medium.
We need notice that this model is just a simplified model because the cavity is not impenetrable in experiment.
In Refs. \cite{michels1937remarks,1938AnP...424...56S,2002JPhB...35..255P,1956CaJPh..34..914H,1984JChPh..80.1569A,1991AmJPh..59..931M,Kang2013},
various methods are introduced to solve a hydrogen atom confined in the impenetrable spherical cavity, such as perturbation method, variational methods,
phase integral method, etc. It is always assumed that the proton in the hydrogen is fixed in the cavity because of the
large mass of proton. In this case, the non-relativistic Hamiltonian of this system is (in atomic unit)
\begin{equation}
H_{atom} = H^0_{atom} +V'(r) ,
\label{eq:hamil2}
\end{equation}
\begin{equation}
H^0_{atom} =-\frac {\nabla ^2}{2}-\frac {1}{r} ,
\label{eq:hamil3}
\end{equation}
\begin{equation}
V'(r)= \begin{cases}
0, & r < r_0; \\
\infty, & r \geq r_0  ,
\label{eq:impener}
\end{cases}
\end{equation}
where $H^0_{atom}$ is the Hamiltonian of the hydrogen in infinite space, $V'(r)$ is the confined potential caused by
the impenetrable spherical cavity and $r_0$ is the radius of the cavity. The proton-electron system (hydrogen) in an
impenetrable spherical cavity with radius $r_0$ is shown in Fig. \ref{fig:hydrogen}.
In Ref. \cite{2019arXiv190205355P},
the authors modified the model to make it closer to the actual situation, where we no longer assume the proton is fixed.
In addition, the modified model can be used to a confined two-body system, whose two constituents have similar mass,
for example quarkonium.
According to Fig. \ref{fig:hydrogen}, the non-relativistic Hamiltonian of a hydrogen atom confined in a cavity is (in atomic unit)
\begin{equation}
H^H_{atom} = -\frac {\nabla_1^2}{2}-\frac {\nabla_2^2}{2\overline{m}_p}-\frac {1}{r_{12}}+V'(r_1)+V'(r_2),
\label{eq:hamil4}
\end{equation}
where $\overline{m}_p$ is the mass of proton, $r_{12}$ is the distance between electron and proton and
$V'(r)$ is given in Eq. (\ref{eq:impener}). However, we need to note that removing the overall kinetic energy of the system is necessary for studying the hydrogen atom in a cavity, similar to the c.m. motion in the infinite space, because the energy spectrum
we calculate is the internal energy of the system. The overall kinetic energy of the system have the same form as the c.m. motion in infinite space, i.e. $-\frac {(\nabla_1+\nabla_2)^2}{2(m_1+m_2)}$. So the corresponding Hamiltonian is modified as:
\begin{eqnarray}
H^{mod}_{atom} = -\frac {\nabla_1^2}{2}-\frac {\nabla_2^2}{2\overline{m}_p}-T_{over}            \notag \\
                 -\frac {1}{r_{12}}+V'(r_1)+V'(r_2)
\label{eq:hamil5}
\end{eqnarray}
with
\begin{equation}
T_{over}=-\frac {(\nabla_1+\nabla_2)^2}{2(1+\overline{m}_p)}.
\label{eq:cmenergy}
\end{equation}
Because the spatial translational invariance of the system is violated, it is meaningless that one separates the motion
of the system into center-of-mass motion and relative motion by introducing the Jacobi coordinates. Otherwise it is
difficult to interpret the boundary conditions of the wave function. For the model of quarkonium confined in an impenetrable
spherical cavity filled with a hot medium, we just need replace the hydrogen atom with the quarkonium and the Coulomb potential
with the Debye screening potential, i.e. $V_{Q \bar Q}(r,T)$ in Eq. (\ref{eq:interquark}). So, the non-relativistic Hamiltonian
of this system is
\begin{eqnarray}
H_{Q\bar Q} =\sum_{i=1}^2 m_i -\frac {\nabla_1^2}{2m_1}-\frac {\nabla_2^2}{2m_2}+
              \frac {(\nabla_1+\nabla_2)^2}{2(m_1+m_2)}            \notag \\
               +V_{Q \bar Q}(r_{12},T)+V'(r_1)+V'(r_2),
\label{eq:hamil6}
\end{eqnarray}
where $r_{12}$ is the distance between quark and antiquark in quarkonium. In this case, the definition of binding energy is
different from the case of infinite space and is not understood well. To calculate the binding energy, we present two limits.
One is the same as the case of infinite space, i.e. $\epsilon^1_i(T)=M_i-2m_{Q}-V_{Q \bar Q}(\infty,T)$. The other assumes
the quarkonium to be dissociated when the quark and antiquark in cavity are pulled to the maximum distance,
i.e $\epsilon^2_i(T)=M_i-2m_{Q}-V_{Q \bar Q}(2r_0,T)$. So the corresponding Hamiltonian of the first limit is
\begin{equation}
H^1_{Q\bar Q}=H^0+V_1(r_{12},T)+V'(r_1)+V'(r_2),
\label{eq:hamil6}
\end{equation}
\begin{equation}
H^0=-\frac {\nabla_1^2}{2m_1}-\frac {\nabla_2^2}{2m_2}+\frac {(\nabla_1+\nabla_2)^2}{2(m_1+m_2)}
\label{eq:hamil0}
\end{equation}
with
\begin{equation}
V_1(r_{12},T) =V_{Q \bar Q}(r_{12},T)-V_{Q \bar Q}(\infty,T).
\label{eq:limit1}
\end{equation}
The corresponding Hamiltonian of the second limit is
\begin{equation}
H^2_{Q\bar Q} =H^0+V_2(r_{12},T)+V'(r_1)+V'(r_2)
\label{eq:hamil6}
\end{equation}
with
\begin{equation}
V_2(r_{12},T) =V_{Q \bar Q}(r_{12},T)-V_{Q \bar Q}(2r_0,T).
\label{eq:limit1}
\end{equation}

\section{Method}\label{sec:method}

In the previous section, we obtain the non-relativistic Hamiltonian for a quarkonium confined in an impenetrable
spherical cavity in two limits mentioned above. The Schr\"{o}dinger equation for the first limit is
\begin{eqnarray*}
	\left[ H_0+V_1(r_{12},T) \right] \Psi^1_i(\boldsymbol{r}_1,\boldsymbol{r}_2)&=&\epsilon^1_i(T) \Psi^1_i(\boldsymbol{r}_1,\boldsymbol{r}_2),    \notag \\
&&\mbox{for}\ r_1,r_2<r_0
\label{eq:schr1}
\end{eqnarray*}
with boundary conditions
\begin{equation}
\Psi^1_i(\boldsymbol{r}_1,\boldsymbol{r}_2)=0,  \mbox{for $r_1 \geq r_0$ or $r_2 \geq r_0$}.
\label{eq:wave1}
\end{equation}
For the second limit, the corresponding Schr\"{o}dinger equation is
\begin{eqnarray}
\left[ H_0+V_2(r_{12},T) \right] \Psi^2_i(\boldsymbol{r}_1,\boldsymbol{r}_2)&=&\epsilon^2_i(T) \Psi^2_i(\boldsymbol{r}_1,\boldsymbol{r}_2), \nonumber \\
&&\mbox{for}\  r_1,r_2<r_0
\label{eq:schr2}
\end{eqnarray}
with the same boundary conditions. $\Delta E^1_i(=-\epsilon^1_i(T))$ and $\Delta E^2_i(=-\epsilon^2_i(T))$ are
the binding energies of state $i$ for the two limits, respectively.
In infinite space, the two-body problem can be reduced to one-body problem by
introducing the center-of-mass motion and relative motion coordinates. However, this procedure does not work for
our problem because the proper boundary condition for the relative motion and the center-of-mass motion is difficult
to establish in our model. So we have to solve the Schr\"{o}dinger equations in independent coordinates $\boldsymbol{r}_1$ and $\boldsymbol{r}_2$.

For the sake of simplicity, the equations in Eq. (\ref{eq:schr1}) and Eq. (\ref{eq:schr2}) are written as $H^H_{Q\bar Q}\Psi_{JM}=\epsilon(T)\Psi_{JM}$, where $\Psi_{JM}$ and $H^H_{Q\bar Q}$ are used to indicate
$\Psi^1_i(\boldsymbol{r}_1,\boldsymbol{r}_2)$ (or $\Psi^2_i(\boldsymbol{r}_1,\boldsymbol{r}_2)$) and
$H^1_{Q\bar Q}$ (or $H^2_{Q\bar Q}$), respectively. Because of the spherical symmetry,
the wavefunction of quarkonium $\Psi_{JM}(\boldsymbol{r}_1,\boldsymbol{r}_2)$ can be written as
$\Psi_{JM}(r_1,r_2,x=cos\Theta)$ (see Fig. \ref{fig:hydrogen}).
In Ref. \cite{2019arXiv190205355P}, we obtain the form of $H_0$, in Eq. (\ref{eq:hamil0}), in coordinates ($r_1,r_2,x$).
So the Hamiltonian in Eqs. (\ref{eq:schr1},\ref{eq:schr2}) can also be written in coordinates ($r_1,r_2,x$).
We can see it is very difficult to obtain the analytic solution of the wavefunction $\Psi_{JM}(r_1,r_2,x)$.
Here we solve the Schr\"{o}dinger equation for $L=0$ states by using the GEM, a powerful various method with high precision \cite{Hiyama:2003cu}. Its reliability on studying the dissociation problem of quarkonium was tested
in Ref. \cite{Cheng:2018pwl}.
We expand the wavefunction $\Psi(r_1,r_2,x)$, i.e. $\Psi_{00}(r_1,r_2,x)$, in terms of a set of basis functions as
\begin{equation}
\Psi(r_1,r_2,x)=\sum_{n=1}^{n_{max}}C_n N_n \Phi_n(r_1,r_2,x),
\label{eq:wave2}
\end{equation}
\begin{equation}
\Phi_n(r_1,r_2,x)=\frac{sin(\frac{\pi r_1}{r_0})sin(\frac{\pi r_2}{r_0})}{r_1 r_2} e^{-\nu_n r_{12}^2}
\label{eq:basiswave}
\end{equation}
with the range parameters taken in geometric progression
\begin{equation}
\nu_n=\frac{1}{b^2_n}, ~~   b_n=b_1 a^{n-1} ~~(n=1,...,n_{max}) ,
\label{eq:range}
\end{equation}
\begin{equation}
a=(\frac {b_{n_{max}}}{b_1})^{1/(n_{max}-1)} .
\label{eq:aaa}
\end{equation}
In Eq. (\ref{eq:wave2}), $N_n$ denotes the normalization constant of the Gaussian basis.
The coefficients $C_n$ of the variational wavefunction in Eq. (\ref{eq:wave2}) are determined by Rayleight-Ritz variational principle.
The Rayleight-Ritz variational principle leads to a generalized matrix eigenvalue problem
\begin{equation}
\sum_{n'=1}^{n_{max}}(H^H_{nn'}-\epsilon(T)N_{nn'})C_{n'}=0 ~~(n=1,...,n_{max}) ,
\label{eq:eigeneq}
\end{equation}
where the energy and overlap matrix elements are given by
\begin{equation}
H^H_{nn'}=\langle \Phi_n \vert H^H_{Q\bar Q} \vert \Phi_{n'} \rangle,
\label{eq:energymatrix}
\end{equation}
\begin{equation}
N_{nn'}=\langle \Phi_n \vert 1 \vert \Phi_{n'} \rangle.
\label{eq:overlapmatrix}
\end{equation}
By solving the eigenvalue problem, we can obtain the coefficients $C_n$, and the corresponding binding
energy $\Delta E(T)(=-\epsilon(T))$. The wavefunction $\Psi(r_1,r_2,x)$ can be obtained by the resulting
coefficients $C_n$. Using the resulting wavefunction, we can obtain the temperature dependence of the average distance according to Eq. (\ref{eq:radii1}).
In this calculation, we set the width of the range parameters and the number of basis functions to be large enough to ensure
the reliability of the calculation: $n_{max}=20$, $b_1=0.1fm$ and $b_{n_{max}}=2r_0$.
\section{Numerical Results}\label{sec:res}

\begin{figure}[!htbp]
\centering
\includegraphics[width=0.36\textwidth]{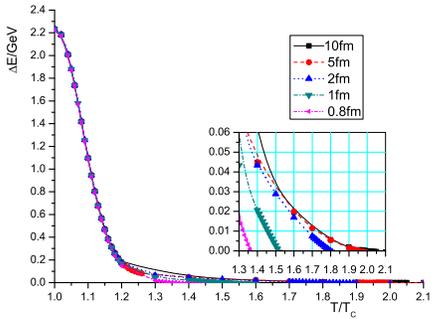}
\includegraphics[width=0.36\textwidth]{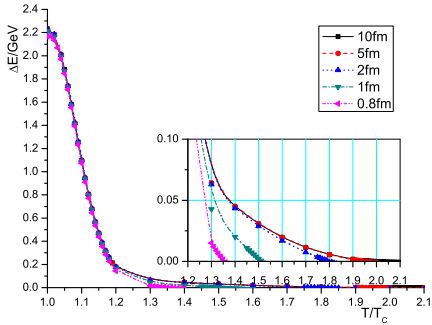}
\caption{\label{fig:bind11}$T$-dependence of binding energy for $J/\psi(1S)$ in two limits:
upper figure for first limit ($V_1$) and lower figure for second limit ($V_2$).}
\end{figure}
\begin{figure}[!htbp]
\centering
\includegraphics[width=0.36\textwidth]{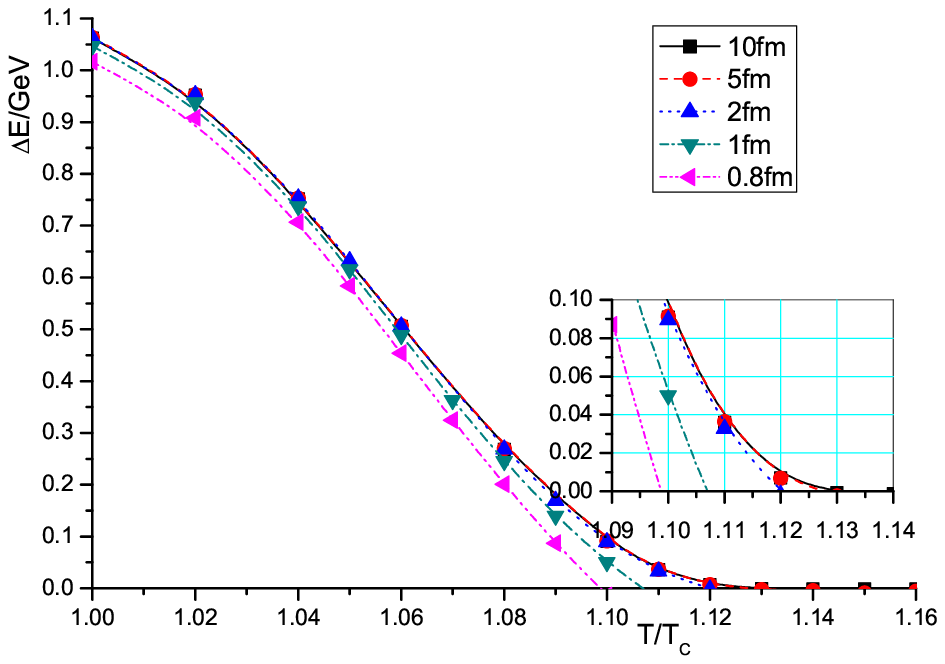}
\includegraphics[width=0.36\textwidth]{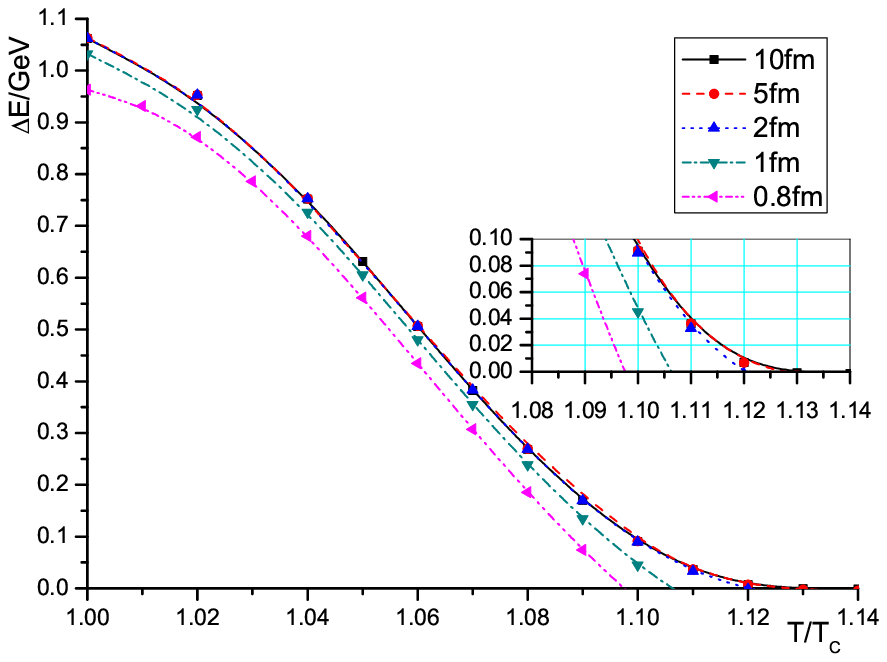}
\caption{\label{fig:bind12}$T$-dependence of binding energy for $\psi'(2S)$ in two limits:
upper figure for first limit ($V_1$) and lower figure for second limit ($V_2$).}
\end{figure}
\begin{figure}[!htbp]
\centering
\includegraphics[width=0.36\textwidth]{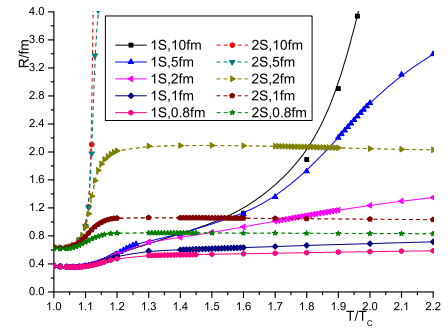}
\includegraphics[width=0.36\textwidth]{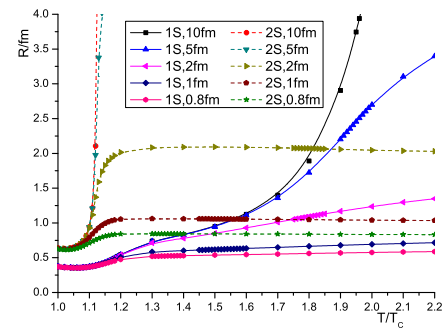}
\caption{\label{fig:radii1}$T$-dependence of average distances for $J/\psi,\psi'$ in two limits:
upper figure for first limit ($V_1$) and lower figure for second limit ($V_2$).}
\end{figure}
In Figs. \ref{fig:bind11},\ref{fig:bind12} and \ref{fig:radii1}, we show the binding energies and
average distances of first two radial states, $1S$ and $2S$, of charmonium in the two limits mentioned above.
\begin{figure}[!htbp]
\centering
\includegraphics[width=0.36\textwidth]{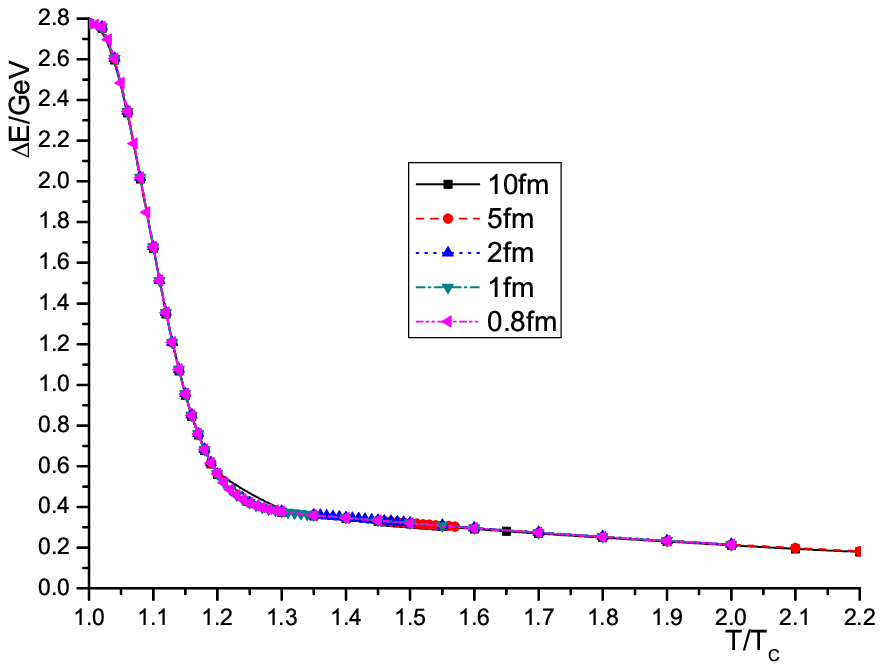}
\includegraphics[width=0.36\textwidth]{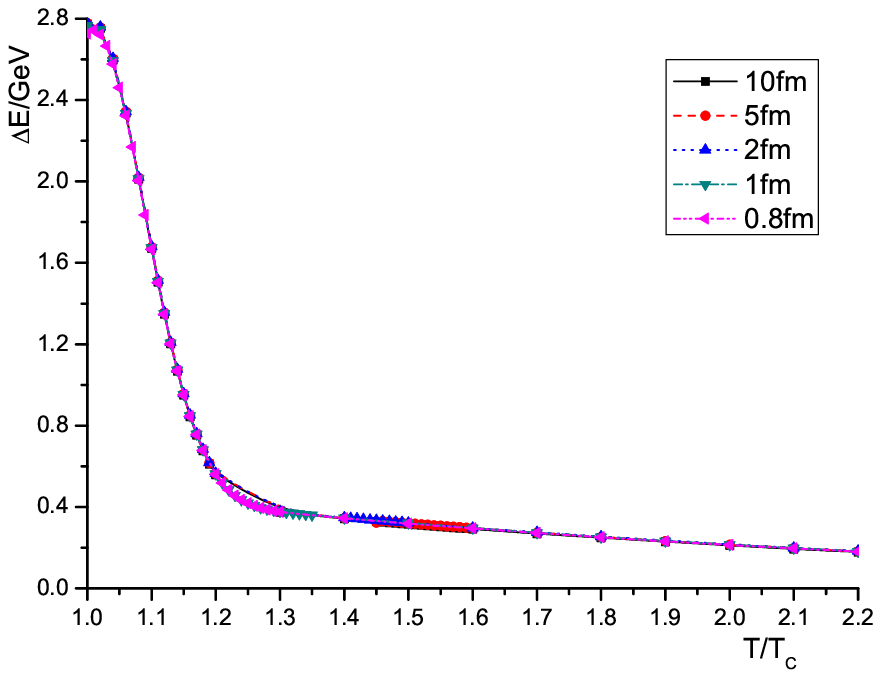}
\caption{\label{fig:bind21}$T$-dependence of binding energy for $\Upsilon(1S)$ in two limits:
upper figure for first limit ($V_1$) and lower figure for second limit ($V_2$).}
\end{figure}
\begin{figure}[!htbp]
\centering
\includegraphics[width=0.36\textwidth]{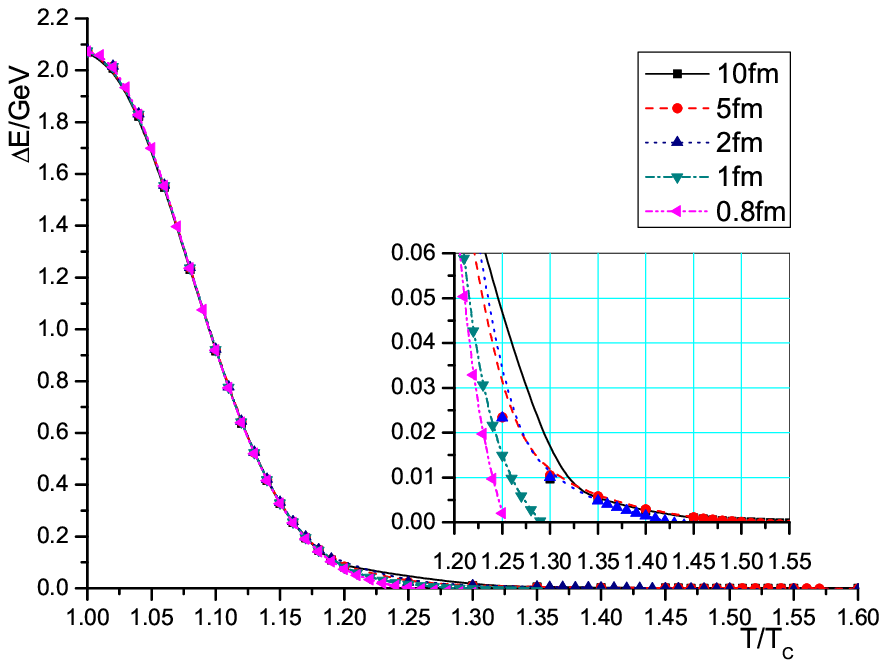}
\includegraphics[width=0.36\textwidth]{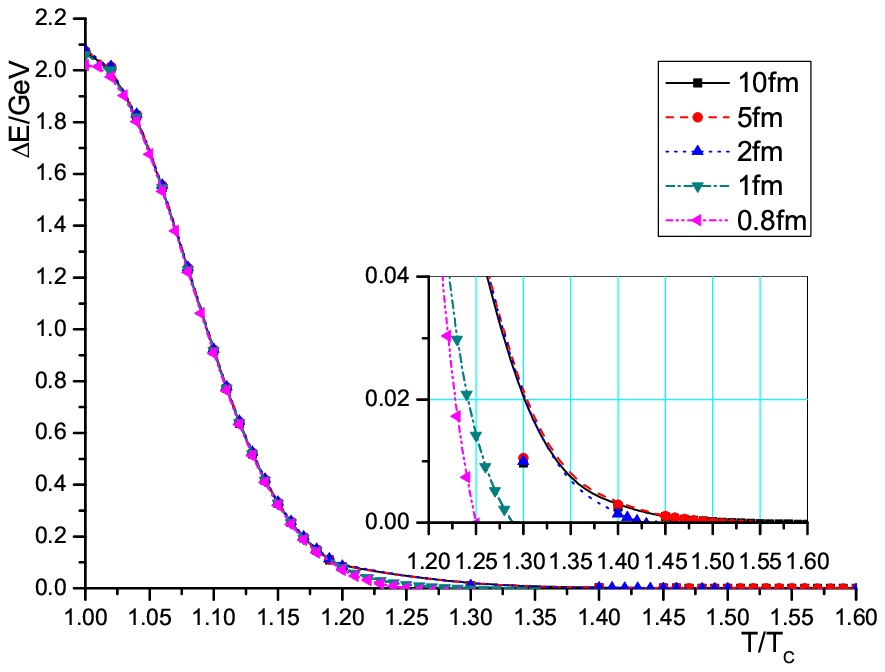}
\caption{\label{fig:bind22}$T$-dependence of binding energy for $\Upsilon'(2S)$ in two limits:
upper figure for first limit ($V_1$) and lower figure for second limit ($V_2$).}
\end{figure}
\begin{figure}[!htbp]
\centering
\includegraphics[width=0.36\textwidth]{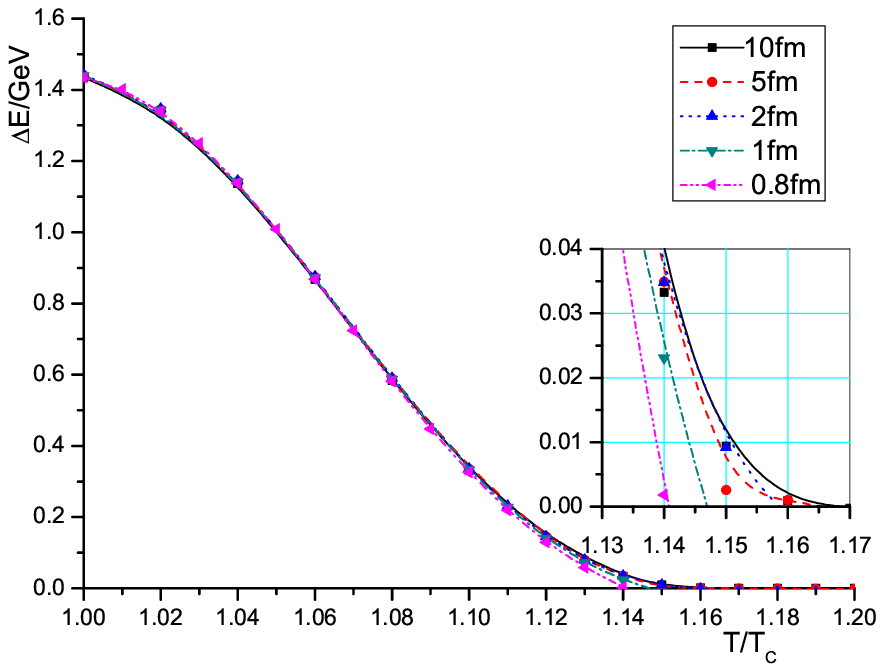}
\includegraphics[width=0.36\textwidth]{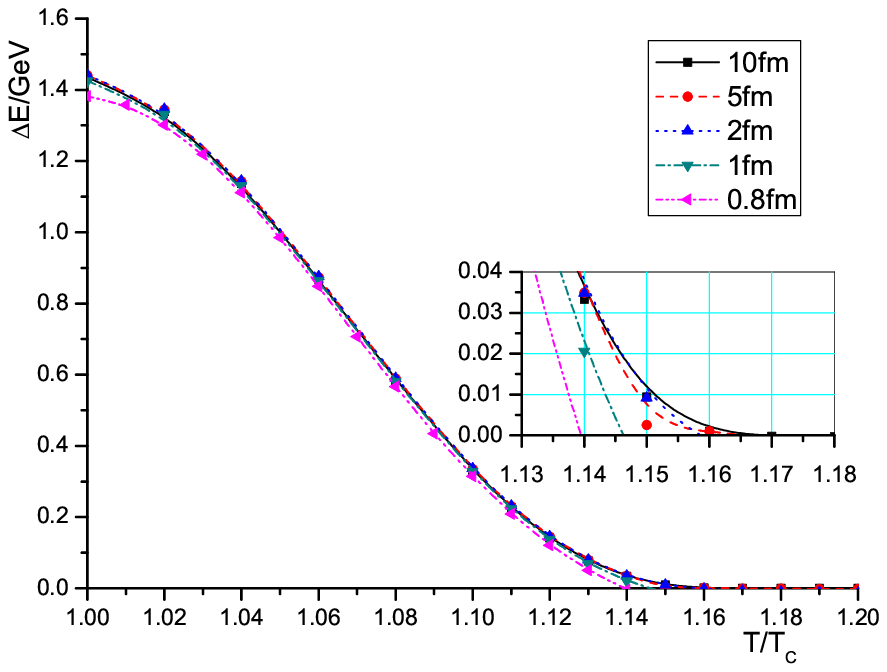}
\caption{\label{fig:bind23}$T$-dependence of binding energy for $\Upsilon''(3S)$ in two limits:
upper figure for first limit ($V_1$) and lower figure for second limit ($V_2$).}
\end{figure}

\begin{figure}[!htbp]
\centering
\includegraphics[width=0.36\textwidth]{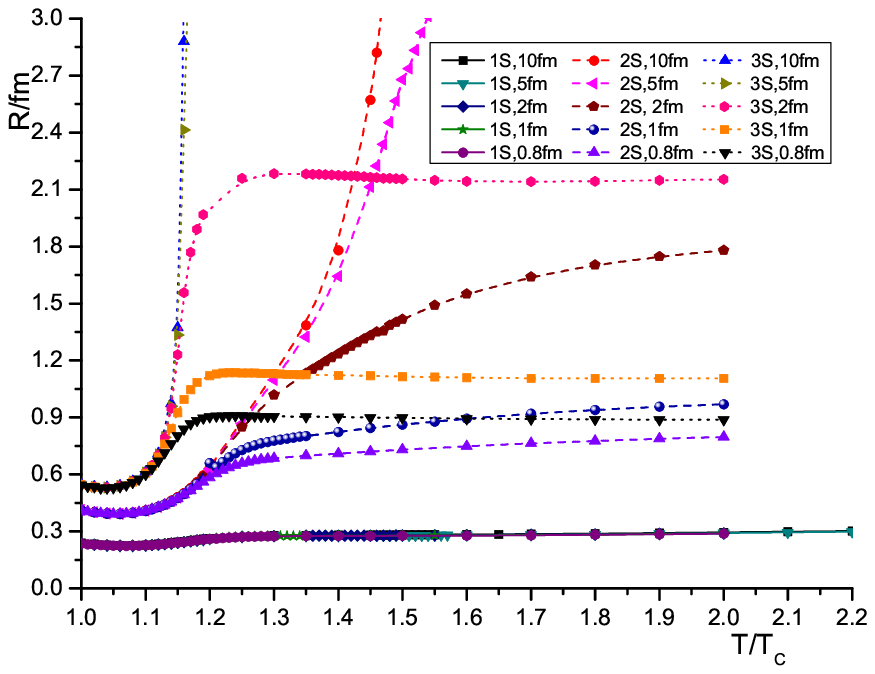}
\includegraphics[width=0.36\textwidth]{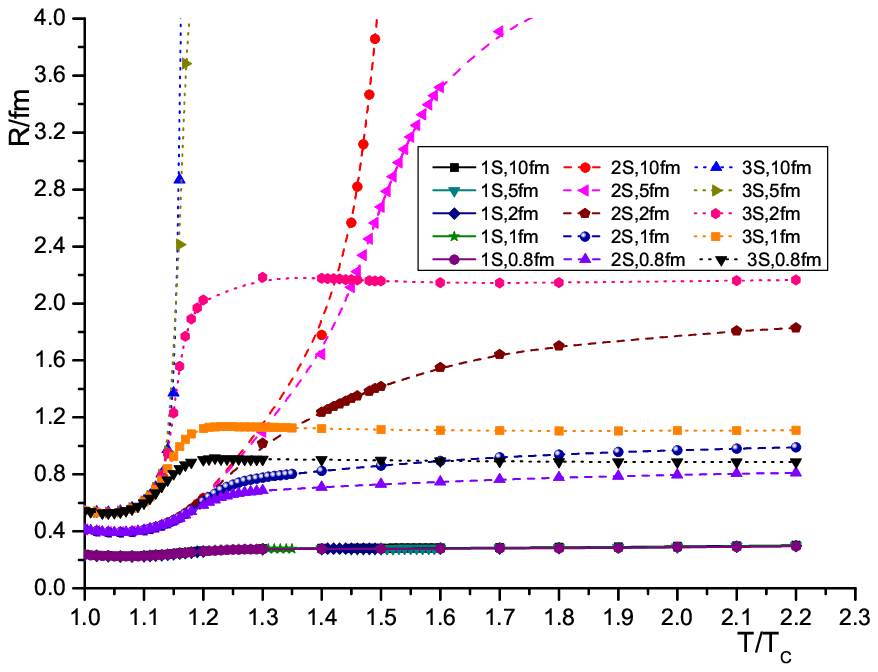}
\caption{\label{fig:radii2}$T$-dependence of average distances for $\Upsilon(1S),\Upsilon'(2S),\Upsilon''(3S)$ in two limits:
upper figure for first limit ($V_1$) and lower figure for second limit ($V_2$).}
\end{figure}

In Figs. \ref{fig:bind21}, \ref{fig:bind22}, \ref{fig:bind23} and \ref{fig:radii2}, we show the binding energies
and average distances of first three radial states, $1S$, $2S$ and $3S$, of bottomonium in the two limits.
Comparing the results between the two limits,
the difference is extremely small. So the two limits we set give almost the same result. In each figure, we show the comparison in the different
sizes of cavity. In Fig. \ref{fig:bind11}, we can see these five lines overlap at low temperature and are separated at high temperature.
The difference among these five lines becomes obvious as the temperature increases, which means the finite volume effects on the results
become larger with the temperature increasing. All of other figures have the same behaviour except Fig. \ref{fig:bind12}.
In Fig. \ref{fig:bind12}, these five lines are separated even at low temperature, which means the finite volume effects on $\psi'$ are
obvious at low temperature. The results in Ref. \cite{Satz:2005hx} shown the $Q\bar Q$ separation distance of $\psi'$ was about $0.9fm$,
very large compared with $J/\psi$. In Figs. \ref{fig:radii1} and \ref{fig:radii2}, we see that the average distances increase as the temperature increases. And we have found the finite volume effects become larger with the temperature increasing.
So we can give the conclusion that the finite volume effects become larger as the size of quarkonium becomes larger.
The $\Upsilon(3S)$ radius is closed to the $\psi'(2S)$ radius, which may lead them to have similar behavior. But here we need to pay attention to the fact that the mass of their constituents are different. The different constituents of $\psi'(2S)$ and $\Upsilon(3S)$ may cause the different behaviours between $\psi'(2S)$ and $\Upsilon(3S)$. There are two parts in the Hamiltonian, the kinetic energy and the potential energy. So the finite volume effects on quarkonium arise from the finite volume effects on these two parts. We can see that the kinetic energy is related with the heavy quark mass. So the
competition between the kinetic energy and potential energy may cause the finite volume effects on $\psi'(2S)$ and $\Upsilon(3S)$ to be different.
To check it, we calculate the average kinetic energy and the average potential energy at different cavity size.
\begin{figure}[!htbp]
\centering
\includegraphics[width=0.36\textwidth]{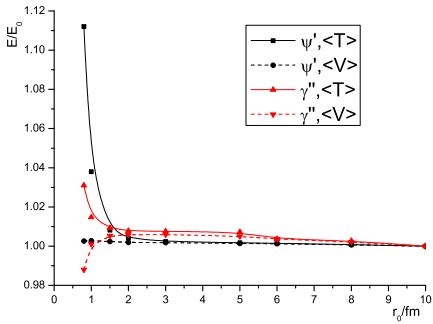}
\caption{\label{fig:TTVV}$r_0$-dependence of the average kinetic energy and the average potential energy for
$\psi'(2S)$ and $\Upsilon(3S)$ at $T_c$.}
\end{figure}
In Fig. \ref{fig:TTVV}, we compare the finite volume effects on the average kinetic energy and the average potential energy between $\psi'(2S)$
and $\Upsilon(3S)$ at $T_c$. $E$ represents the average kinetic energy or the average potential energy, and $E_0$ is the corresponding
energy value at $r_0=10fm$. We can see that the average kinetic energy of $\psi'(2S)$ is very sensitive to the cavity size,
but the average potential energy is not. And the average kinetic energy and average potential energy of $\Upsilon(3S)$
are not very sensitive to the cavity size. Comparing the behaviours between $\psi'(2S)$ and $\Upsilon(3S)$,
it is found that, with the increase in the mass of $c$ quark to $b$ quark, the average kinetic energy become less sensitive to the cavity sizes.
Meanwhile, for $\psi'(2S)$ and $\Upsilon(3S)$, the average potential energy are not very sensitive to the cavity size.
As a result, the $\psi'(2S)$ binding energy is sensitive to the cavity size but $\Upsilon(3S)$ binding energy is not near $T_c$.
So we can give a conclusion that the less mass of the constituents of $\psi'(2S)$ ($c$ quark) leads the $\psi'(2S)$ binding energy to
be sensitive to the cavity size but $\Upsilon(3S)$ binding energy to be not at low temperature.

In Figs. \ref{fig:bind11} and \ref{fig:bind12}, we show the resulting binding energy behaviour for different charmonium states
in two limits and different sizes
of cavity, respectively. When they vanish, the bound states no longer exist, so that $\Delta E(T)=0$ determines the dissociation temperature. The results for
dissociation temperatures of charmonium in Ref. \cite{Cheng:2018pwl} and our calculation results are listed in Table \ref{tab:ccdisso}.
\begin{table}[!h]
\begin{center}
\caption{\label{tab:ccdisso} Dissociation temperatures $T_d/T_c$ of charmonium in different sizes of cavity and two limits.}
\renewcommand\arraystretch{1.8}
\begin{tabular}{cccc}
\hline
 potential & $r_0$[fm] & $1S$ & $2S$  \\
\hline
 & $\infty$ (Ref. \cite{Cheng:2018pwl}) & 2.06 & 1.13 \\
\hline
$V_1$ & 10 & 2.01 & 1.13 \\
\hline
 & 5 & 1.96 & 1.13 \\
\hline
 & 2 & 1.8 & 1.12 \\
\hline
 & 1 & 1.52 & 1.107 \\
\hline
 & 0.8 & 1.36 & 1.098 \\
\hline
$V_2$ & 10 & 2.01 & 1.13 \\
\hline
 & 5 & 1.96 & 1.13 \\
\hline
 & 2 & 1.8 & 1.12 \\
\hline
 & 1 & 1.52 & 1.107 \\
\hline
 & 0.8 & 1.36 & 1.097 \\
\hline
\end{tabular}
\end{center}
\end{table}
In Figs. \ref{fig:bind21}, \ref{fig:bind22} and \ref{fig:bind23}, we show the resulting binding energy behaviour
for different bottomonium states in two limits and different sizes of cavity, respectively. The results for
dissociation temperatures of bottomonium in Ref. \cite{Cheng:2018pwl} and our calculation results are listed in
Table \ref{tab:bbdisso}.
\begin{table}[!h]
\begin{center}
\caption{\label{tab:bbdisso} Dissociation temperatures $T_d/T_c$ of bottomonium in different size of cavity and two limits.}
\renewcommand\arraystretch{1.8}
\begin{tabular}{ccccc}
\hline
 potential & $R_0$[fm] & $1S$ & $2S$ & $3S$  \\
\hline
 & $\infty$ (Ref. \cite{Cheng:2018pwl}) & 5.81 & 1.56 & 1.17 \\
\hline
$V1$ & 10 & $>$2.2 & 1.52 & 1.17 \\
\hline
 & 5 & $>$2.2 & 1.52 & 1.17 \\
\hline
 & 2 & $>$2.2 & 1.43 & 1.16 \\
\hline
 & 1 & $>$2.2 & 1.3 & 1.147 \\
\hline
 & 0.8 & $>$2.2 & 1.26 & 1.142 \\
\hline
$V2$ &10 & $>$2.2 & 1.52 & 1.17 \\
\hline
 & 5 & $>$2.2 & 1.51 & 1.17 \\
\hline
 & 2 & $>$2.2 & 1.43 & 1.16 \\
\hline
 & 1 & $>$2.2 & 1.29 & 1.147 \\
\hline
 & 0.8 & $>$2.2 & 1.25 & 1.139 \\
\hline
\end{tabular}
\end{center}
\end{table}
Due to the free energy of quark-antiquark system just fitting the lattice date from $0.8T_c$ to $2T_c$,
we only show $T_d$ of $\Upsilon(1S)$ is $> 2.2T_c$. There have been many lattice improvements since Ref. \cite{Digal:2005ht}.
For example, Ref. \cite{Bazavov:2018wmo} calculates the free energy up to $T \sim 2 GeV$. We may obtain an
approximate value of the $\Upsilon(1S)$ dissociation temperature based on this work in a future work.
We can see the dissociation temperatures of charmonium and bottomonium decrease with the radius of cavity decreasing.
For charmonium, the changes in the dissociation temperature of $J/\psi(1S)$ is more obvious compared with that of
$\psi'(2S)$. For bottomonium, the changes in the dissociation temperature of $\Upsilon'(2S)$ is more obvious compared
with that of $\Upsilon''(3S)$.  At $r_0 \geq 5fm$, the size of the cavity is much larger than that of the quakronium.
In this case, the quarkonium can be seen as being in infinite space. It can explain why the changes in resulting
dissociation temperatures of each states is negligible at $r_0 \geq 5fm$.
From Figs. \ref{fig:radii1} and \ref{fig:radii2}, we can see the radius reaches a finite value when the temperature is
higher than dissociation temperature, which is different from infinite space. This is because the quarkonium is confined
in an impenetrable cavity. After the ground state (or excited state) of quarkonium dissociating, the resulting quark
and antiquark are bounded in the impenetrable cavity. So the distance between the quark and antiquark is finite and
less than the diameter of the corresponding cavity. With the cavity radius decreasing, the finite value decreases.

\section{discussion and Conclusions}\label{sec:summ}
In infinite space, the free energy of quark-antiquark system we construct based on Debye-H\"uckel theory
fits the lattice data quite well for all $r$ from $0.8T_c$ to $2T_c$. For the system of hydrogen in an impenetrable spherical cavity, the model we proposed is different from other people and closer to the reality. In this model, separating the motion of the system into center-of-mass motion and relative motion by introducing the Jacobi coodinates is meaningless and we solve the equation using independent coordinates $\boldsymbol{r}_1$ and $\boldsymbol{r}_2$.
From the temperature dependence of binding energy and average distance for charmonium and bottomonium, we can see that they have almost the same behaviour at low temperature but the different behaviour at high temperature except for $\psi'$, which means the finite volume effects on these states are negligible at low temperature and become more obvious with temperature increasing. The state $\psi'$ has different behaviours of the binding energy at different cavity radius even at low temperature. Compared the behaviours among $J/\psi,\psi',\psi''$, we can give a conclusion that the bigger radius of the quarkonium leads the finite volume effect to become more obvious. Compared the behaviours between $\Upsilon(3S)$ and $\psi'$, we can give a conclusion that the less mass of the constituents leads the finite volume effect to become more obvious.

The results on the dissociation temperatures of quarkonium show that the dissociation temperatures decrease with the cavity radius decreasing. Compared with the state $\psi'(2S)$ and $\Upsilon''(3S)$, the changes in the states $J/\psi(1S)$ and $\Upsilon'(2S)$ are more obvious. Because of the quarkonium confined in an impenetrable cavity, the average distance increases to a finite value rather than an infinite value.
The behaviour of the quarkonium confined in an impenetrable cavity is very different from that in infinite space.

It should be pointed out that the model we present is a simplified model and the real fireball produce in heavy ion collisions
is not impenetrable.
The reason why we consider a impenetrable cavity is that it is difficult for us to
solve the non-infinitely deep potential well if we take a non-impenetrable cavity into account.
In addition, there are some effects, arising from magnetic field, finite baryon density and so on, contributing
to quarkonium dissociation. Such work deserves our progressive consideration.

\section*{Acknowledgement}

This work is supported in part by the National Natural Science Foundation of China (under Grants Nos. 11475085, 11535005, 11690030)
and by Nation Major State Basic Research and Development of China (2016YFE0129300). And X.Luo is supported by the National Natural
Science Foundation of China (under Grants Nos.11575069, 11828501, 11890711 and 11861131009) and Fundamental Research Funds for the Central
Universities (No.CCNU19QN054).

\bibliography{refs}

\end{document}